\documentclass{ws-ijmpcs}
\usepackage{braket,psfrag}
\usepackage{bm}

\def\trimmarks{}

\newcommand{\vecc}[1]{\mbox{\boldmath $#1$}}

\begin{document}

\markboth{Cherednikov \& Stefanis}
{TMD at the edge of the lightcone}


%
\catchline{}{}{}{}{}
%

\title{TRANSVERSE-MOMENTUM-DEPENDENT \\
       PARTON DISTRIBUTIONS
       AT THE EDGE OF THE LIGHTCONE\footnote{Invited talk presented at Workshop ``QCD evolution of parton distributions: from collinear to non-collinear case'', 8 - 9 Apr 2011, Thomas Jefferson National Accelerator Facility, Newport News (VA), USA. Preprint \tt{RUB-TPII-03/2011}.}}

\author{I. O. CHEREDNIKOV$^\dag$\footnote{On leave of absence from
        Joint Institute for Nuclear Research, BLTP JINR,
        RU-141980 Dubna, Russia}}

\address{Departement Fysica, Universiteit Antwerpen, \\
         B-2020 Antwerpen, Belgium\\
$^\dag$E-mail: igor.cherednikov@ua.ac.be}

\author{N. G. STEFANIS$^\ddag$}

\address{Institut f\"{u}r Theoretische Physik II, \\
         Ruhr-Universit\"{a}t Bochum,
         D-44780 Bochum, Germany\\
$^\ddag$E-mail: stefanis@tp2.ruhr-uni-bochum.de}

\maketitle


\begin{abstract}
We present a completely gauge-invariant operator definition
of transverse-momentum-dependent parton densities (TMD),
supplied with longitudinal lightlike gauge links as well as
transverse gauge links at lightcone infinity.
Within this framework, we consider the consistent treatment of
specific divergences, emerging in the ``unsubtracted'' TMD beyond
the tree approximation, and construct the soft factors to cancel
unphysical singularities.
We confront this approach with factorization schemes, which make
use of covariant gauges with off-the-lightcone gauge links, and
discuss their mutual connection.
\end{abstract}


\section{Introduction}
Different operator definitions of the transverse-momentum-dependent
parton distribution functions (PDF)---TMD for short in what
follows\cite{Sop77,Col78,RS79,CS81,CS82,CSS85}---are actively
discussed in the literature, see, e.g.,
Refs.~\refcite{BJY03,BMP03,Hau07,CM04,CKS10,Lat_TMD,Col03,BR05,CRS07,Col08,AR11,Col11}
and references therein.
Among the principal issues to be addressed in any consistent operator
definition of the TMD, there are their gauge invariance, the
cancelation of undesirable divergences, renormalization-group
properties, and ultraviolet (UV) and rapidity evolution equations.
Moreover, properly defined TMDs have to be incorporated into the QCD
factorization formula for the structure functions of semi-inclusive
processes that can be schematically written as
\begin{equation}
  F_{\rm semi-incl.}
\sim
  {H } \otimes {{\cal F}_D}
  \otimes {{\cal F}_F} \otimes {S} \ ,
\label{eq:factor_symb}
\end{equation}
where $H$ is the hard (perturbatively calculable) part,
${\cal F}_{D, F}$ are the TMD distribution and/or fragmentation
functions, and $S$ is the soft factor---a specific ingredient in the
semi-inclusive factorization approach.
Nonperturbative distribution functions of partons
(in what follows we consider only quark distributions),
depending on the longitudinal components $x$, as well as on the
transverse components $\vecc k_\perp$, of their momenta accumulate
information about the {\it intrinsic} motion of the hadron's
constituents.
Initially, the TMDs have been considered as a direct generalization
of the collinear PDFs:
\begin{equation}
 {F}_{\rm [coll.]} (x, \mu)
\ \to \
 {\cal F}_{\rm [tmd]} (x, \mathbf k_\perp, \mu, \eta) \ , \
 \int\! d^2 k_\perp \ {\cal F}_{\rm [tmd]}(x, \mathbf k_\perp, \mu, \eta)
\to
 { {F}_{\rm [coll.]} (x, \mu) } \ .
\label{eq:tmd_soper}
\end{equation}
However, such a straightforward relationship between TMDs and integrated
PDFs can only be proved in the tree approximation because rapidity
divergences jeopardize this procedure, or even entail its breakdown.
Note that in the collinear case the only scale is the UV renormalization
parameter $\mu$, while in the TMD case, an additional dependence from the
rapidity cutoff $\zeta$ arises\cite{Sop77,CS81,CS82}.

In this work, we present and discuss an operator definition of the quark
TMD that embodies gauge invariance in terms of lightlike longitudinal and
transverse gauge links with the appropriate behavior at lightcone
infinity\cite{CKS10,CS_all}.

\section{Gauge links in the lightcone gauge}

We start with the ``unsubtracted'' definition (i.e., without the
explicit isolation of the soft factor) of the ``quark in a quark'' TMD
(where the subscript on $[{\rm A}]_{\rm n}$ denotes the axial lightcone
gauge for lightlike longitudinal gauge links along the vector $n$,
employing the notations in Ref.~\refcite{CheISMD}):
\begin{eqnarray}
&& {\cal F}_{\rm unsubtr.}^{[{\rm A}_{\rm n}]} \left(x, {\mathbf k}_\perp; \mu, \eta\right)
=
  \frac{1}{2}
  \int \frac{d\xi^- d^2\bm{\xi}_\perp}{2\pi (2\pi)^2}
  {\rm e}^{-ik \cdot \xi}
  \left\langle
              p |\bar \psi_i (\xi^-, \bm{\xi}_\perp)
              [\xi^-, \bm{\xi}_\perp;
   \infty^-, \bm{\xi}_\perp]_{n}^\dagger  \right.  \nonumber \\
   && \left.
\times
   [\infty^-, \bm{\xi}_\perp;
   \infty^-, \bm{\infty}_\perp]_{\vecc l}^\dagger
   \gamma^+[\infty^-, \bm{\infty}_\perp;
   \infty^-, \mathbf{0}_\perp]_{\vecc l}
   [\infty^-, \mathbf{0}_\perp; 0^-,\mathbf{0}_\perp]_{n}
   \psi_i (0^-,\mathbf{0}_\perp) | p
   \right\rangle \
\label{eq:general}
\end{eqnarray}
with ${\xi^+=0}$.
The path-ordered longitudinal (lightlike, $n^2=0$) $[...]_{n}$
and transverse $[...]_{\vecc l}$ gauge links
\begin{eqnarray}
\begin{split}
& [\infty^-, \mbox{\boldmath$\xi_\perp$}; \xi^-, \mbox{\boldmath$\xi_\perp$}]_{n}
\equiv {}
  {\cal P} \exp \left[
                      - i g \int_0^\infty d\tau \ n_{\mu}^- \
                      A_{a}^{\mu}t^{a} (\xi + n^- \tau)
                \right] \, , \\
&
  [\infty^-, \mbox{\boldmath$\infty_\perp$};
  \infty^-, \mbox{\boldmath$\xi_\perp$}]_{\vecc l}
\equiv {}
  {\cal P} \exp \left[
                      - i g \int_0^\infty d\tau \ \mbox{\boldmath$l$}
                      \cdot \mbox{\boldmath$A$}_{a} t^{a}
                      (\mbox{\boldmath$\xi_\perp$}
                      + \mbox{\boldmath$l$}\tau)
                \right] \,
\end{split}
\end{eqnarray}
ensure the formal gauge invariance of this TMD.
The TMD can be normalized as follows:
\begin{eqnarray}
&& {\cal F}_{\rm unsubtr.}^{[{\rm A}_{\rm n}] (0)} (x, {\mathbf k}_\perp)
=
  \frac{1}{2}
  \int \frac{d\xi^- d^2\bm{\xi}_\perp}{2\pi (2\pi)^2}
  {\rm e}^{- i k^+ \xi^- + i \bm{k}_\perp \cdot \bm{\xi}_\perp}
\nonumber \\
&&
   \times { \langle p | }\bar \psi (\xi^-, \bm{\xi}_\perp)
   \gamma^+
   \psi (0^-,0_\perp) {| p \rangle } =
   \delta(1 - x ) \delta^{(2)} (\vecc k_\perp) \ .
\label{eq:tree_tmd}
\end{eqnarray}
In the tree approximation (which we distinguish from
${\cal F}_{\rm unsubtr.}^{[{\rm A}_{\rm n}] (0)}$ by keeping
``classical'' gauge links in the former)
straightforward integration over the transverse momentum
$\vecc k_\perp$ immediately yields the collinear gauge-invariant PDF
\begin{eqnarray}
  & & \int\! d^2 k_\perp \ {\cal F}_{\rm unsubtr.}^{[{\rm A}_{\rm n}],\rm tree} (x, {\mathbf k}_\perp)
  =
  {F}_{\rm [coll.]}^{\rm tree} (x) \ , \nonumber \\
  & & {F}_{\rm [coll.]}^{\rm tree} (x)
=
  \frac{1}{2}
  \int \frac{d\xi^- }{2\pi }
  {\rm e}^{-ik^{+}\xi^{-} }
  \left\langle
              p |\bar \psi_i (\xi^-, \mbox{\boldmath$0_\perp$})[\xi^-, 0^-]_n
              \gamma^+
   \psi_i (0^-,\mbox{\boldmath$0_\perp$}) | p
   \right\rangle \ .
\end{eqnarray}
Let us emphasize that the off-the-light-cone operator definitions
(A$_{\rm v}$-, C$_{\rm v}$-TMD given in Ref.~\refcite{CheISMD}) do not
satisfy this relation---even in the tree approximation.
Instead, they can be shown to obey a factorized formula in terms of
collinear PDFs at small $\vecc b_\perp$ in the impact-parameter
representation\cite{Col11}.
The use of one of these approaches is a matter of convenience, provided
that the corresponding operator definition of the TMD is consistent
with the factorization scheme, cf. Eq. (\ref{eq:factor_symb}).
However, one has to be careful when comparing different classes of
definitions, like
A$_{\rm v}$-, C$_{\rm v}$-TMD vs. A$_{\rm n}$-, C$_{\rm n}$-TMD,
since these are, in principle, different objects even at the tree
level.
In contrast, the comparison of the definitions
A$_{\rm n}$ vs. C$_{\rm n}$ is justified, since they represent the
same quantities in different gauges.

\begin{figure}
\centering
\includegraphics[scale=0.5,angle=90]{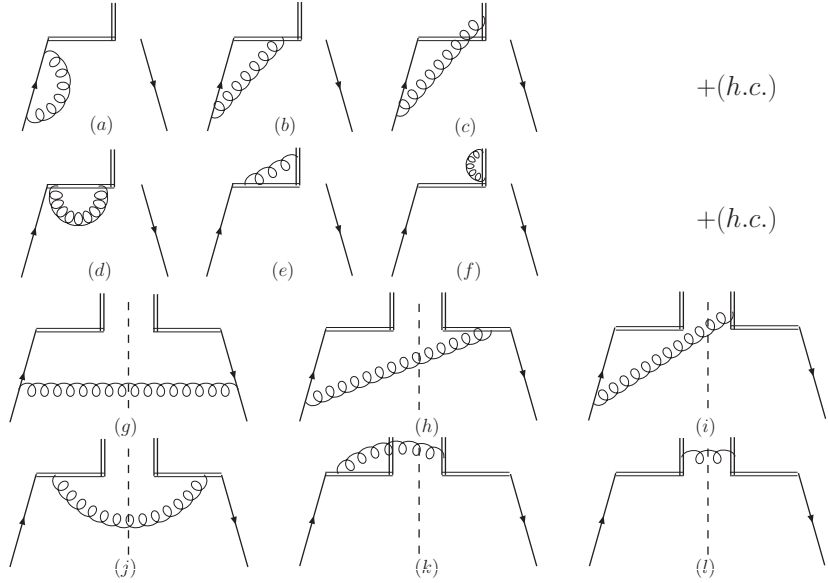}~~
\caption{Complete set of the one-loop diagrams corresponding to the
gauge-invariant quark TMD PDF without soft-term contributions.
\label{fig:graphs}}
\end{figure}

Beyond the tree level, the quark fields in the operator definition of
the TMD (\ref{eq:general}) have to be considered as Heisenberg field
operators, i.e.,
\begin{equation}
 \psi_i(\xi)
=
  {\rm e}^{- ig\left[ \int\! d\eta \ \bar \psi \hat {\cal A} \psi \right]} \
  \psi_i^{\rm free} (\xi)
  \ , \
  \left[ \int\! d\eta \ \bar \psi \hat {\cal A} \psi \right]\
  \equiv
  \int\! d^4\eta \ \bar \psi (\eta) \gamma_\mu
  \psi (\eta) {\cal A}^\mu (\eta) \ .
\label{eq:fermion_heis}
\end{equation}
Therefore, the $\mathcal{O}(g)$ contributions from the gauge links are
contracted with the quark-gluon interaction terms
$\left[ \int\! d\eta \ \bar \psi \hat {\cal A} \psi \right]$,
originating from the Heisenberg fields (\ref{eq:fermion_heis}), and
give rise to the set of the $\mathcal{O}(g^2)$ one-gluon exchange
graphs shown in Fig.~\ref{fig:graphs}
(more details are given in Ref.\ \refcite{SCK10}).

In the light-cone gauge, some of these diagrams disappear.
Let us show that the contributions of the longitudinal (lightlike)
gauge links (which are shown in Fig.~\ref{fig:graphs} by horizontal
double lines, whereas the transverse gauge links are denoted by vertical
double-lines) cancel in the lightcone gauge $A^+=0$ when combined with
$\eta$-dependent pole prescriptions for the gluon propagator.
Formally, the ``classical'' exponential $\exp[- ig \int\! dz^- A^+]$
equals unity in the lightcone gauge, irrespective of the applied pole
prescription, but this is worth being proven for the prescriptions we
use in our analysis.

Next, we evaluate the gauge field, the source of which is a charged
pointlike particle (a struck quark) moving with the quasi-constant
four-velocity $v_\mu$ along the straight line $v_\mu \tau$.
The corresponding ``classical'' current then is
\begin{equation}
  J_\mu (y)
=
  g \int\! d y'_\mu \ \delta^{(4)} (y - y') \ , \quad
  y'_\mu = v_\mu \ \tau \ .
\label{eq:current-2}
\end{equation}
The velocity changes its direction only at the origin, where the
collision with the hard photon takes place and the quark deviates
from its initial ``trajectory''.
The gauge field defined by such a current then reads
\begin{equation}
  A_{\rm LC}^\mu (\xi)
=
  \int\! d^4 y \ {\cal  D}^{\mu\nu}_{\rm LC}(\xi - y) J_\nu (y)\ \  ,
\label{eq:source1}
\end{equation}
where ${\cal D}^{\mu\nu}_{\rm LC}$ is the free gluon propagator in the
lightcone gauge.
This gauge field is exactly what forms the longitudinal gauge
link\cite{BJY03}.
We assume that the velocity of the struck quark is parallel to the
``plus''- and the ``minus''- lightcone vectors $n^{\pm}$ before
and after the hard collision, respectively:
\begin{eqnarray}
  J_\mu (y)
& = &
    g \left[ n^+_\mu \int_{-\infty}^0 \! d\tau \ \delta^{(4)}(y - n^+ \tau)
  + n^-_\mu  \int_{0}^\infty \! d\tau \ \delta^{(4)}(y - n^+ \tau)\right]
\nonumber \\
& = &
    g \ \delta^{(2)} (\vecc y_\perp) \left[ n^+_\mu \delta(y^-)  \int\!
    \frac{dq^-}{2\pi} \frac{{\rm e}^{-iq^- y^+}}{q^- + i0}
    - n^-_\mu \delta(y^+)  \int\! \frac{dq^+}{2\pi}
    \frac{{\rm e}^{-iq^+ y^-}}{q^+ - i0}\right] \ .
\label{eq:current-3}
\end{eqnarray}
Then, one has
\begin{equation}
  A_{\rm LC}^\mu (\xi)
=
  - g \ n_\nu^+ \int\! \frac{d^4 q}{2(2\pi)^4}
   \ {{\rm e}^{- i q \cdot \xi}} \ \tilde {\cal D}^{\mu \nu}_{\rm LC} (q)
   \int\! dy^+ dy^- d^2 y_\perp {{\rm e}^{ i q \cdot y}} \delta(y^-)
                                \delta^{(2)} (\vecc y_\perp)
 \ ,
\label{eq:perp_sour1}
\end{equation}
where the free gluon propagator in the lightcone gauge
$A^+ = (A \cdot n^-)=0$ reads
\begin{equation}
  {\cal D}^{\mu\nu}_{\rm LC} (z)
=
   \! \int\! \frac{d^4 q}{(2\pi)^4} \
   {{\rm e}^{- i q \cdot z}} \tilde {\cal D}^{\mu\nu}(q)
= - \! \int\! \frac{d^4 q}{(2\pi)^4} \
    \frac{{\rm e}^{- i q \cdot z}} {
    q^2+i0}
    \left( g^{\mu\nu}
  - \frac{q^{\mu}(n^-)^{\nu}+q^{\nu}(n^-)^{\mu}}{[q^+]_\eta}
    \right)
\label{eq:gluon-prop}
\end{equation}
and the pole prescription $[q^+]_\eta$ has yet to be defined.
We neglect the quark and gluon masses, since we are mainly interested in the
UV and rapidity singularities.

After the integration over the variable $y$, we get
\begin{equation}
  A_{\rm LC}^\mu (\xi)
=
  - g \ n_\nu^+ \int\! \frac{d^4 q}{2(2\pi)^4}
   \ {{\rm e}^{- i q \cdot \xi}} \ \frac{\delta(q^-)}{q^2+ i0} \
   \left( g^{\mu\nu}
  - \frac{q^{\mu}(n^-)^{\nu}+q^{\nu}(n^-)^{\mu}}{[q^+]_\eta}
    \right) \ .
\label{eq:delta}
\end{equation}
Now we are able to calculate the plus-component of the gauge
field (\ref{eq:delta}) $A^+ = (A \cdot n^-)$:
\begin{equation}
  A_{\rm LC}^\mu
=
  - \frac{1}{2} g \ \int\! \frac{d^4 q}{(2\pi)^4}
   \ {{\rm e}^{- i q \cdot \xi}} \ \frac{\delta(q^-)}{q^2+ i0} \left[ n^{+\mu}
  - \frac{q^{\mu}}{[q^+]_\eta}
    \right] \ .
\label{eq:gauge_f_1}
\end{equation}
The first integral is trivial and yields
(using dimensional regularization $\omega = 4 - 2 \epsilon$ for the
integration over transverse degrees of freedom)
\begin{equation}
   \int\! \frac{d^\omega q}{(2\pi)^\omega}
   \ {{\rm e}^{- i q \cdot \xi}} \frac{\delta(q^-)}{q^2+ i0}
   =
   \frac{1}{2\pi} \delta(\xi^-) \ \int\! \frac{d^{\omega-2} q}{(2\pi)^{\omega-2}}
   \ \frac{{\rm e}^{i \bm{q}_\perp \cdot \bm{\xi}_\perp}}{- \vecc q_\perp^2+i0} \ .
\label{eq:gauge_0}
\end{equation}
To perform the $\eta$-dependent integral, we observe that
\begin{equation}
  \int\! \frac{d^\omega q}{(2\pi)^\omega}\
  \frac{{\rm e}^{- i q \cdot \xi}}{q^2 + i0} \ \frac{q^+}{[q^+]_\eta} \ \delta(q^-)
=
  i \frac{\partial}{\partial \xi^-} \
  \int\! \frac{d^\omega q}{(2\pi)^\omega} \
  \frac{{\rm e}^{- i q \cdot \xi}}{q^2 + i0}\frac{\delta(q^-)}{[q^+]_\eta}  \  \ .
\label{eq:gauge_f_2}
\end{equation}
Making use of $\delta(q^-)$ in Eq. (\ref{eq:gauge_f_2}), we separate
out the transverse part
\begin{equation}
  \int\! \frac{d^\omega q}{(2\pi)^\omega} \
  \frac{{\rm e}^{- i q \cdot \xi}}{q^2 + i0} \ \frac{1}{[q^+]_\eta} \ \delta(q^-)
=
  \frac{1}{2\pi} \ \int\! \frac{dq^+}{2\pi} \
  \frac{{\rm e}^{- i q^+\xi^-}}{[q^+]_\eta} \ \int\! \frac{d^{\omega-2} q}{(2\pi)^{\omega-2}}
   \ \frac{{\rm e}^{i \bm{q}_\perp \cdot \bm{\xi}_\perp}}{- \vecc q_\perp^2+i0} \ .
\label{eq:gauge_f_3}
\end{equation}
Consider now the longitudinal integral which we define as
\begin{equation}
  \int\! \frac{dq^+}{2\pi} \ \frac{{\rm e}^{- i q^+\xi^-}}{[q^+]_\eta}
=
  \lim_{\eta \to 0} \
  \int\! \frac{dq^+}{2\pi} \ \frac{{\rm e}^{- i q^+\xi^-}}{q^+ \pm i \eta} \ ,
\label{eq:gauge_f_4}
\end{equation}
where $\pm$ corresponds, respectively, to the retarded and advanced
prescription, while the principal-value prescription can be obtained by
symmetrization.
Integral (\ref{eq:gauge_f_4}) can be evaluated using the residue
theorem to read
\begin{equation}
  \int\! \frac{dq^+}{2\pi} \ \frac{{\rm e}^{- i q^+\xi^-}}{q^+ \pm i \eta}
=
  \mp i \theta(\pm \xi^-)  {\rm e}^{\mp \xi^- \eta} \ , \quad
  \theta (z) = \left\{ {1 \ , \ z > 0 \atop 0 \ , \ z < 0}\right.
  \ .
\label{eq:gauge_f_5}
\end{equation}
Taking the derivative with respect to $\xi^-$ and employing
Eq. (\ref{eq:gauge_f_2}), we find in the limit $\eta \to 0$ the
expression
\begin{equation}
  \lim_{\eta \to 0} i \frac{\partial}{\partial \xi^-} \
  \int\! \frac{d^\omega q}{(2\pi)^\omega} \
  \frac{{\rm e}^{- i q \xi}}{q^2 + i0}\frac{\delta(q^-)}{[q^+]_\eta}
=
  \frac{1}{2\pi} \  \delta(\xi^-) \int\! \frac{d^{\omega-2} q}{(2\pi)^{\omega-2}}
  \ \frac{{\rm e}^{i \bm{q}_\perp \cdot \bm{\xi}_\perp}}{- \vecc q_\perp^2+i0} \ .
\label{eq:gauge_6}
\end{equation}
Reversing the sign and adding this result to Eq. (\ref{eq:gauge_0}), we
get, according to Eq. ({\ref{eq:gauge_f_1}}),
\begin{equation}
  A^+ = (A \cdot n^-)
=
  0 \ .
\label{eq:gauge_final_0}
\end{equation}
The ``minus''-component of the gauge field
\begin{equation}
  A^-
=
  (A^\mu \cdot n^+_\mu)
  \sim
  n^+_\mu
    \left( (n^+)^\mu
  - \frac{q^{\mu} + q^- (n^-)^{\mu}}{[q^+]}
    \right)
=
    0 - \frac{2 q^-}{[q^+]}
\end{equation}
vanishes as well after carrying out the integration over $q^-$
in Eq. ({\ref{eq:delta}}) by virtue of
$\delta(q^-)$, i.e.,
$
  A^-
=
  0
$.

This suffices to show that the use of the $\eta$-dependent pole
prescription in the gluon propagator is consistent with the
lightcone gauge $A^+ = 0$, while for a further discussion we refer
to Refs.~\refcite{Col11,CDL84}.
A formal proof of a similar statement concerning the
Mandelstam-Leibbrandt pole prescription will be presented separately.

\section{Lightcone TMD: One-loop effects}

In the previous section, we have shown that in the lightcone gauge
with $\eta$-dependent pole prescriptions for the gluon propagator,
the longitudinal (lightlike) gauge links are equal to unity:
$\exp[- ig \int\! dz^- A^+] = 1$.
Therefore, the diagrams Fig.~\ref{fig:graphs} $(b, d, e, f, h, j, k)$
(and their Hermitian conjugate parts---if any) give zero contributions
in the lightcone gauge, despite the opposite claims by Collins
in Ref.~\refcite{Col11}.
Moreover, we give below a formal justification of our statements by
demonstrating the gauge invariance of our framework.

In the lightcone gauge, the $\eta$-dependent contribution of diagram
$1(a)$ is
\begin{equation}
  {\Sigma}^{(a)} [{\rm A}_{\rm n}]
=
  - \frac{\alpha_s}{\pi} C_{\rm F} \  \Gamma(\epsilon)\
  \left[ 4 \pi \frac{\mu^2}{-p^2} \right]^\epsilon\
  \delta (1-x) \delta^{(2)} (\bm{k}_\perp)\ \int_0^1\!
  dx \frac{(1-x)^{1-\epsilon}}{x^\epsilon [x]_\eta} \ .
\label{eq:sigma-lc}
\end{equation}
In a covariant gauge (definition ${\rm C}_{\rm n}$), the counterpart of
this contribution stems from  the vertex diagram $1(b)$ and reads
\begin{equation}
 \Sigma^{(c)}[{\rm C}_{\rm n}]
=
 - \frac{\alpha_s}{\pi} C_{\rm F}
 \Gamma (\epsilon) \left[ 4\pi \frac{\mu^2}{-p^2} \right]^{\epsilon} \
 \delta (1-x) \delta^{(2)} (\bm{k}_\perp )\
 \int_0^1\! dx \ \frac{x^{1-\epsilon}}{(1-x)^{1+\epsilon}} \ .
\label{eq:sigma-cov}
\end{equation}
By a trivial change of variables and by using the
$\eta$-regularization, the singular integral above can be written as
\begin{equation}
  {\Sigma}^{(c)} [{\rm C}_{\rm n}]
=
  - \frac{\alpha_s}{\pi} C_{\rm F} \  \Gamma(\epsilon)
  \left[ 4 \pi \frac{\mu^2}{-p^2} \right]^\epsilon
  \delta (1-x) \delta^{(2)} (\bm{k}_\perp) \int_0^1\!
  dx \frac{(1-x)^{1-\epsilon}}{x^\epsilon (x \pm i \eta)}
=
  {\Sigma}^{(a)} [{\rm A}_{\rm n}] \ .
\label{eq:sigma-lc_cov}
\end{equation}
This is an interesting result and supports the validity of the operator
definition of the TMD with the lightlike longitudinal gauge links,
justifying its gauge invariance at the one-loop level.
Analogous results have been obtained recently within the framework of
soft collinear effective theory (SCET)\cite{IS10,GIS11,LMP11}.
Similar problems have been addressed and resolved long ago in
Ref.~\refcite{Ste83}.

Therefore, whatever gauge is adopted, the one-loop corrections to the
``unsubtracted'' definition (\ref{eq:general}) give rise to
pathological {\it overlapping divergences} that comprise UV and
rapidity poles simultaneously
$
  {\sim \frac{1}{\varepsilon} \ \ln \eta }
$.
This jeopardizes the renormalizability of TMDs and calls for a
certain {generalized} renormalization procedure to augment the
insufficient dimensional regularization.
In our works\cite{CS_all}, we worked out such a procedure that
enabled us to obtain a well-defined and fully gauge-invariant TMD PDF,
free of undesirable divergences.
This TMD PDF has calculable renormalization-group properties and obeys
an evolution equation with respect to rapidity in the impact parameter
space.
Moreover, the one-loop analysis of the UV anomalous dimension of the
``unsubtracted'' TMD PDF in (\ref{eq:general}) with the
$\eta$-dependent pole prescription in the lightcone gauge shows that
the overlapping singularities produces a correction to the anomalous
dimension that can be identified at this order with the well-known
{\it cusp anomalous dimension}\cite{KR87}.
The upshot of this discussion is that in order to renormalize the
``naive'' TMD PDF, given by (\ref{eq:general}), one has to introduce
an additional renormalization factor, which depends on
$\bar \eta = \eta/p^+$ and can be written as the vacuum average of the
gauge links evaluated along a special contour with an obstruction
(cusp)---see Ref.~\refcite{CKS10,CS_all} for more details:
\begin{equation}
  Z_{\eta}^{-1}
=
  \left\langle 0 \left| {\cal P} \exp \Bigg[ ig \int_{\chi}d\zeta^\mu
  \ t^a A^a_\mu (\zeta) \Bigg] \right| 0 \right\rangle \ .
\end{equation}

\begin{figure}[h]
  \includegraphics[scale=0.57,angle=90]{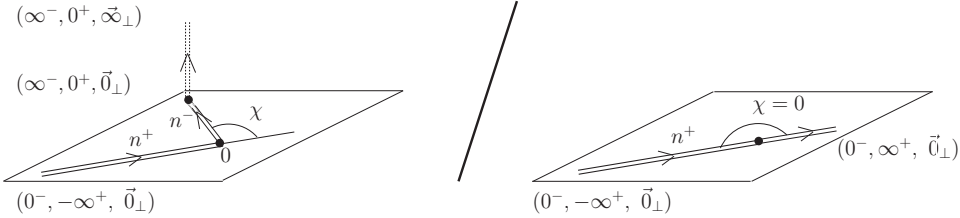}
  \caption{Subtraction of the infinite self-energy contribution of the
  lightlike gauge links in the soft factor.
  \label{fig:soft-factor}}
\end{figure}

However, the overlapping singularities are not the only ones that have
to be removed from the proper definition of the TMD.
The one-loop corrections to the soft factor itself give rise to another
type of unphysical divergences.
For instance, in the lightcone gauge, the self-energy contribution of
the soft factor still contains an uncanceled singularity of the term
\begin{equation}
  \Sigma_{\rm soft}[{\rm A}_{\rm n}]
=
  i g^2 \mu^{2\epsilon} C_{\rm F}2 p^+ \ \int\! \frac{d^\omega q}{(2\pi)^\omega}
  \frac{1}{q^2 (q^- \cdot p^+ - i0) [q^+]_\eta} \ ,
\label{eq:soft}
\end{equation}
which appears to be rapidity-independent\cite{CS_all}.
This calls for an additional subtraction of this self-energy part
that is presented graphically in Fig.\ \ref{fig:soft-factor}.
This subtraction does not affect the rapidity evolution equations and
does not break the factorization structure (\ref{eq:factor_symb}).
In fact, it has an intuitively clear physical interpretation: it serves
to remove unobservable contributions due to the self-energy of the
infinite lightlike gauge links that are mere artifacts of the
unobservable background.

The definition of the ${\rm A}_{\rm n}$ TMD PDF, which embodies the
above requirements, reads\cite{CS_all}
\begin{eqnarray}
&& {\cal F}_{\rm i/h}^{[{\rm A}_{\rm n}]} \left(x, {\mathbf k}_\perp; \mu, \eta\right)
=
  \frac{1}{2}
  \int \frac{d\xi^- d^2\bm{\xi}_\perp}{2\pi (2\pi)^2}
  {\rm e}^{-ik^+ \cdot \xi^- + i \bm{k}_\perp \bm{\xi}_\perp} \cdot \nonumber \\
&&  \times \left\langle
              h |\bar \psi_i (\xi^-, \bm{\xi}_\perp)
              [\xi^-, \bm{\xi}_\perp;
   \infty^-, \bm{\xi}_\perp]_{n}^\dagger  [\infty^-, \bm{\xi}_\perp;
   \infty^-, \bm{\infty}_\perp]_{\vecc l}^\dagger \right. \ \gamma^+ \cdot \nonumber \\
   && \left.
\times
   [\infty^-, \bm{\infty}_\perp;
   \infty^-, \mathbf{0}_\perp]_{\vecc l}
   [\infty^-, \mathbf{0}_\perp; 0^-,\mathbf{0}_\perp]_{n}
   \psi_i (0^-,\mathbf{0}_\perp) | h
   \right\rangle
   \cdot R^{-1} \ ,
\label{eq:final_tmd}
\end{eqnarray}
with the soft factor
\begin{eqnarray}
&& R^{-1}(\mu, \eta)
= \nonumber \\
&& \frac{\langle 0
  |   \ {\cal P}
  \exp\Big[ig \int_{\mathcal{C}_{\rm cusp}}\! d\zeta^\mu
           \ {\cal A}^\mu (\zeta)
      \Big] \cdot
  {\cal P}^{-1}
  \exp\Big[- ig \int_{\mathcal{C'}_{\rm cusp}}\! d\zeta^\mu
           \ {\cal A}^\mu (\xi + \zeta)
      \Big]
  {| 0
  \rangle } }
{ \langle 0
  |   \ {\cal P}
  \exp\Big[ig \int_{\mathcal{C}_{\rm smooth}}\! d\zeta^\mu
           \ {\cal A}^\mu (\zeta)
      \Big] \cdot
  {\cal P}^{-1}
  \exp\Big[- ig \int_{\mathcal{C'}_{\rm smooth}}\! d\zeta^\mu
           \ {\cal A}^\mu (\xi + \zeta)
      \Big]
  | 0
  \rangle  } \ ,
\nonumber
\label{eq:soft_factor}
\end{eqnarray}
where ${\cal A} \equiv t^a A^a$ and the contours for the soft factor
are displayed in Fig.\ \ref{fig:soft-factor}.

\section{Light-cone TMD: Evolution equations}

After subtracting the soft factor, the operator definition (\ref{eq:final_tmd}) is multiplicatively renormalizable and obeys the following
one-loop evolution equation with respect to the UV scale
$\mu$:
\begin{equation}
  \mu \frac{d }{d\mu} \ {\cal F}^{[{\rm A}_{\rm n}]} \left(x, {\mathbf k}_\perp; \mu, \eta\right)
=
  (\gamma_{\rm LC}
  + O(\alpha_s^2) ) \  {\cal F}^{[{\rm A}_{\rm n}]}  \left(x, {\mathbf k}_\perp; \mu, \eta\right)\ ,
\label{eq:uv_nr}
\end{equation}
where
$\gamma_{\rm LC} = \frac{4 \alpha_s}{3 \pi} C_{\rm F}$
is the anomalous dimension of the bilocal quark operator in
the lightcone gauge.
It is interesting to note that, in the lightcone gauge with the
Mandelstam-Leibbrandt prescription, the ``unsubtracted'' TMD
(\ref{eq:general}) has an anomalous dimension that is free
of any undesirable contributions and reads
\begin{eqnarray}
& & \mu \frac{d }{d\mu} \ \left[ {\cal F}_{\rm unsubtr.}^{[{\rm A}_{\rm n}^{\rm ML}]} \cdot R_n^{-1} \right]
=
  \mu \frac{d }{d\mu} \  {\cal F}_{\rm unsubtr.}^{[{\rm A}_{\rm n}^{\rm ML}]} \nonumber \\
  & & =
    (\gamma_{\rm LC}+ O(\alpha_s^2)) \  \left[ {\cal F}_{\rm unsubtr.}^{[{\rm A}_{\rm n}^{\rm ML}]} \cdot R_n^{-1} \right]
=
    ( \gamma_{\rm LC} + O(\alpha_s^2)) \ {\cal F}_{\rm unsubtr.}^{[{\rm A}_{\rm n}^{\rm ML}]}\ .
\label{eq:uv_ml}
\end{eqnarray}
In the one-loop approximation, these results are in agreement with
those given in Ref.~\refcite{AR11}.

In order to study the {\it rapidity} evolution, we have to concentrate
on the specific TMD singularities which only depend on the additional
rapidity parameter $\eta$, but else do not violate the
renormalizability of the TMDs.
These singular terms have to be resummed by means of an equation of
the Collins--Soper type\cite{CS81,CS82}.
Our framework indeed allows the development of such a resummation
procedure of the rapidity divergences as we now show.

To this end, recall that the $\eta$-dependent part (\ref{eq:sigma-lc})
of the quark self-energy diagram corresponding to Fig. 1$(a)$ has the
form
\begin{eqnarray}
  {\Sigma}^{(a)} [{\rm A}_{\rm n}]
& = &
  - \frac{\alpha_s}{\pi} C_{\rm F} \  \Gamma(\epsilon)\
  \left[ 4 \pi \frac{\mu^2}{-p^2} \right]^\epsilon\
  \delta (1-x) \delta^{(2)} ({\boldmath k_\perp})\ \nonumber \\
& \times &
  \frac{1}{p^+} \left[- 1 - \ln\frac{\pm i \eta}{p^+}
  - \epsilon \left( 2 - \frac{\pi^2}{3}
  - \frac{1}{2} \ln^2 \frac{\pm i \eta}{p^+}  \right) + O(\epsilon^2) \right] \ ,
\label{eq:sigma-lc_rapidity}
\end{eqnarray}
where the quadratic term $\sim \ln^2 \eta$ is responsible for the
``rapidity'' evolution.
The linear terms stem from the remaining virtual and real graphs of the
``unsubtracted'' TMD and the soft factor.

In order to confront our framework with the Collins-Soper rapidity
evolution approach, we make use of the impact representation of the TMD
\begin{equation}
  {\cal F} \left(x, {\mathbf b}_\perp; \mu, \eta\right)
  =
  \int\! d^2 {\mathbf k_\perp} \ {\rm e}^{i \bm{b}_\perp \cdot \bm{k}_\perp}\
  {\cal F} \left(x, {\mathbf k}_\perp; \mu, \eta\right) \ .
\end{equation}
Then, the Collins-Soper rapidity evolution equation
(which holds for the off-the-light-cone
A$_{\rm v}$-TMD and C$_{\rm v}$-TMD) becomes
\begin{equation}
  \zeta \frac{\partial }{\partial \zeta} \ {\cal F}_{[{\rm A,C}_{\rm v}]} \left(x, {\mathbf k}_\perp; \mu, \zeta\right)
=
  \left[ K_{\rm v} (\mu, b_\perp) + G_{\rm v} (\mu, \zeta) \right] \
  {\cal F}_{[{\rm A,C}_{\rm v}]} \left(x, {\mathbf k}_\perp; \mu, \zeta\right)
  \ ,
\label{eq_cs_axial}
\end{equation}
where the functions $K$ and $G$ have the following
renormalization-group properties:
\begin{equation}
  \mu \frac{d}{d\mu} K_{\rm v}
=
  - \mu \frac{d}{d\mu} G_{\rm v} =
  \gamma_{\rm cusp} \ .
\label{eq:kg_cusp}
\end{equation}
In our approach, the analogue of the Collins-Soper rapidity cutoff
is given by the new variable
$\theta \equiv (p\cdot n)/\eta$.
Therefore, the corresponding evolution equation takes the form
\begin{equation}
  \theta \frac{\partial }{\partial \theta} \
  {\cal F}_{[{\rm A}_{\rm n}]} \left(x, {\mathbf k}_\perp; \mu, \theta \right)
=
  \left[ K_{\rm n} (\mu, \vecc b_\perp) + G_{\rm n} (\mu, \theta) \right] \
  {\cal F}_{[{\rm A}_{\rm n}]} \left(x, {\mathbf k}_\perp; \mu, \theta\right)
  \ ,
\label{eq_cs_lc}
\end{equation}
noting that the limit $\eta \to 0$ corresponds to the limit
$\zeta \to \infty$ in the Collins-Soper approach.
Here the sum $K_{\rm n} + G_{\rm n}$ for the A$_{\rm n}$-TMD can be
evaluated perturbatively in the small-$\vecc b_\perp$ region.
The relation (\ref{eq:kg_cusp}) for the A$_{\rm n}$-TMD has been
verified at the one-loop level in our previous works\cite{CS_all}.
Explicit results for the rapidity evolution kernel
$\left[ K_{\rm n} (\mu, \vecc b_\perp) + G_{\rm n} (\mu, \theta) \right]$
will be reported in the future.

The bottom line is: the generalized definition of the TMD PDF
(28) allows one to derive its renormalization-group
evolution with respect to the UV scale $\mu$ and to resum the large
rapidity logarithms by means of the evolution equation
(\ref{eq_cs_lc}), formally akin to that in the Collins-Soper
procedure, designed for off-the-light-cone quantities.

\section{Conclusions and discussion}

To conclude, we have shown that the completely gauge-invariant
operator definition of the transverse-momentum-dependent parton
densities, which makes use of longitudinal lightlike gauge
links and also of appropriate transverse gauge links at lightcone
infinity, can be consistently formulated beyond the tree-level
approximation---at least in the one-loop order.
The advantages of the presented approach are:
\begin{itemize}
\item
In the lightcone axial gauge, the longitudinal gauge links cancel,
while the transverse gauge links can be eliminated by proper boundary
conditions for the gauge fields at lightcone infinity.
This provides a natural framework for QCD factorization schemes, where
the use of covariant gauges and not purely lightlike gauge links is not
convenient.
\item
A direct relationship between the (unintegrated) TMDs and the
(integrated) collinear PDFs can be established by straightforward
$\vecc k_\perp$-integration.
One ultimately gets the well-known collinear and gauge-invariant PDF
that fulfills the DGLAP evolution equation (in the one-loop order).
\item
The transverse gauge links at lightcone infinity (supplemented by
appropriate boundary conditions) accumulate information about the
initial- and/or final-state interactions in the axial gauges, in which
the longitudinal gauge links disappear.
The transverse gauge links naturally shrink to harmless constants
after performing the $\vecc k_\perp$-integration without causing a
breakdown of the standard structure of the longitudinal gauge links
in the collinear PDF, while in the case of the off-the-light-cone
definitions this is, at least, not obvious.
\item
Moreover, the transverse gauge links are responsible for $T$-odd
effects in the lightcone gauge that makes it possible to apply this
framework to the study of phenomenologically important quantities,
such as the Sivers, Boer-Mulders, Collins functions, etc.
\item
A complete proof of the QCD factorization within the presented
scheme---in particular, the evaluation of the hard part in the
lightcone gauge and the explicit demonstration of the independence
of the full structure functions of unphysical scales is still lacking
and will be treated in the future.
\end{itemize}


\end{document}